\documentclass[3p,times,preprint]{elsarticle}

\usepackage{ecrc}

\volume{00}

\firstpage{1}

\journalname{Nuclear Physics A}

\runauth{A. Ohnishi et al.}

\jid{NUPHA}

\jnltitlelogo{Nuclear Physics A}

\CopyrightLine{2012}{Published by Elsevier Ltd.}

\usepackage{amssymb}
\usepackage[figuresright]{rotating}

\newcommand{\LL}{$\Lambda\Lambda$}
\newcommand{\DBLL}{\Delta B_{\Lambda\Lambda}}
\newcommand{\reff}{r_\mathrm{eff}}
\newcommand{\fm}{\mathrm{fm}}
\newcommand{\MeV}{\mathrm{MeV}}
\newcommand{\Fig}[2]{\includegraphics[width=#1]{#2}}
\usepackage{amsmath}

\begin{document}

\begin{frontmatter}

\dochead{}



\title{Exotic hadrons and hadron-hadron interactions\\ 
in heavy ion collisions\tnoteref{report}}
\tnotetext[report]{
Talk presented at
the 11th International Conference on Hypernuclear and Strange Particle Physics (HYP2012), Oct.1-5, 2012, Barcelona, Spain.
Report No.: YITP-13-5
}

\author[a]{A. Ohnishi}
\author[b]{S. Cho}
\author[c]{T. Furumoto}
\author[d]{T. Hyodo}
\author[a]{D. Jido}
\author[e]{C. M. Ko}
\author[a]{K. Morita}
\author[b]{S. H. Lee}
\author[f]{M. Nielsen}
\author[g]{T. Sekihara}
\author[g]{S. Yasui}
\author[h]{K. Yazaki}
\author[]{(ExHIC Collaboration)}
%\author[a,b]{Author3}

\address[a]{Yukawa Institute for Theoretical Physics, Kyoto University, Japan}
\address[b]{Institute of Physics and Applied Physics, Yonsei University, Korea}
\address[c]{Ichinoseki National College of Technology, Japan}
\address[d]{Department of Physics, Tokyo Institute of Technology, Japan}
\address[e]{Cyclotron Institute and Department of Physics and Astronomy,
Texas A \& M University, USA}
\address[f]{Instituto de F{\'i}sica, Universidade de S{\~a}o Paulo, Brazil}
\address[g]{Institute of Particle and Nuclear Studies, KEK, Japan}
\address[h]{RIKEN Nishina Center, Japan}

\begin{abstract}
%bla, bla, bla
We discuss the exotic hadron structure and hadron-hadron interactions
in view of heavy ion collisions.
First, we demonstrate that
a hadronic molecule with a large spatial size
would be produced more abundantly in the coalescence model
compared with the statistical model result.
Secondly, we constrain the $\Lambda\Lambda$ interaction
by using the recently measured $\Lambda\Lambda$ correlation data.
We find that the RHIC-STAR data favor 
the \LL\ scattering parameters in the range
$1/a_0 \leq -0.8~\fm^{-1}$ and $\reff \geq 3~\fm$.
\end{abstract}

\begin{keyword}
Exotic hadrons 
\sep
Heavy ion collisions
\sep
Hadronic molecule
\sep
Two particle correlation
\sep
$\Lambda\Lambda$ interaction
\sep
H particle
\end{keyword}

\end{frontmatter}

% \linenumbers

\section{Introduction}
\label{Sec:Intro}

The first dozen of years in the 21st century
may be recognized as the starting point
of the exotic hadron renaissance.
In textbooks, hadrons are explained as
$\bar{q}q$ (mesons) or $qqq$ (baryons) composites,
and many of the hadron masses are well described
in the quark model.
This common understanding of hadrons becomes doubtful
in these years.
Starting from $D_{sJ}(2317)$~\cite{Babar-DsJ},
we have found many hadronic states which we cannot
understand in the na\"ive quark model.
For example, a penta quark state $udud\bar{s}$
is claimed to be observed
at LEPS~\cite{LEPS-Penta},
while its existence is still controversial~\cite{CLAS-Penta}.
$Z^+(4430)$ is a typical and clear exotic hadron~\cite{Belle-Z}:
Its mass is close to $D_1\bar{D}^*$ threshold
and we expect it contains $\bar{c}c$,
and it has a positive charge.
Thus the minimum quark content
of $Z^+(4430)$ is $\bar{c}c\bar{d}u$.

Understanding the structure of exotic hadrons is important
in order to construct a new scheme over the quark model
to categorize hadrons including normal and exotic ones.
There are mainly two-types of structure considered for exotic hadrons.
One of them is compact multi-quark structure,
and the other is hadronic molecule structure.
These two types of structure would have different sizes.
We expect that a multi-quark state has a similar size to normal hadrons,
while the deuteron, a well-known hadronic molecule, has a much larger size
than normal hadrons.

Another aspect of exotic hadron physics is that
it is related to the hadron-hadron interaction.
The existence of $\Lambda(1405)$ below the $\bar{K}N$ threshold
leads to various aspects of $\bar{K}N$ interaction.
Similarly, once the pole position of the $S=-2$ dibaryon ($H$) is fixed,
$\Lambda\Lambda$ interaction is strongly constrained.

High energy heavy ion collisions would provide unique information
on exotic hadron structure and hadron-hadron interactions.
Various hadrons are produced abundantly in heavy ion collisions,
and it is natural to expect that exotic hadrons are also produced.
The dynamics of high energy heavy ion collisions is so complex
that statistical argument becomes valid; 
We may dare to say, heavy-ion collisions are simple and clean.
Via the hadron-hadron correlation measurement,
it is in principle possible to extract the resonance pole
above the strong decay threshold.

In this proceedings,
we discuss the exotic hadron structure and hadron-hadron interaction
in view of heavy-ion collisions.
In Sec.~\ref{Sec:ExHIC}, we demonstrate that the production yield
is sensitive to hadron size.
In Sec.~\ref{Sec:LL}, we discuss \LL\ correlation and its relation
to \LL\ interaction.

\section{Exotic Hadron Yields in Heavy Ion Collisions}
\label{Sec:ExHIC}

Several mechanisms have been proposed so far to gain energy in exotic states.
Based on the diquark picture and color-magnetic interaction,
exotic hadron states including heavy-quarks should exist in some channels.
The color-magnetic interaction is proportional to $1/m_i/m_j$,
where $m_i$ is the quark mass.
Then in a state made of $\bar{Q}\bar{Q}ud$ ($Q$ denotes a heavy quark),
a diquark $(ud)(\bar{Q}\bar{Q})$ component would be favored
rather than a mesonic molecule component $(\bar{Q}u)(\bar{Q}d)$.
$T^1_{cc}(J^\pi=1^+,I=0)$ is one of theoretically proposed
hadronic states made of $\bar{c}\bar{c}ud$~\cite{Tcc}.
Since the strong decay to $DD$ is forbidden
by the angular momentum conservation
and its predicted mass is lower than the $D_1D$,
$T^1_{cc}$ may have a small width.
A hadronic molecular state with the same quantum number
is also predicted in the pion-exchange model~\cite{OPE-Tcc}.
The pion couples $D(0^-)$ and $D^*(1^-)$ states
and pseudo-scalar and vector meson masses are close with heavy-quarks
(heavy-quark symmetry),
then the two "states" ($D^*D$ and $DD^*$) couple strongly
and $T^1_{cc}$ can gain energy.

% Size
It is a challenge to clarify the structure and mechanism
for each of the exotic states to exist.
One of the key quantity to distinguish
a multi-quark state and a hadronic molecule state is the size:
When it is a multi-quark state such as a diquark pair $(ud)(\bar{Q}\bar{Q})$,
the confining force between the diquarks would make the system compact.
When it is a hadronic molecule such as a $D^*D$ bound state,
the exotic hadron size would be determined by its binding energy
or the range of the pion-exchange.
Recent work on $\Lambda(1405)$ 
%by Sekihara, Hyodo, and Jido
sheds light on this idea~\cite{Sekihara}.
They have proposed that the evidence of the $\bar{K}N$ picture
of $\Lambda(1405)$ may be found in the negatively large squared charge radius,
which may be observed in the electric form factor.
%The measurement of the electric form factor of short-lived hadrons
%is not an easy task.

We have proposed that we can utilize high-energy heavy-ion collisions
to obtain knowledges on the exotic hadron size~\cite{ExHIC}.
At RHIC and LHC, abundant hadrons are produced
and their yield ratio is well described by the statistical model,
which assumes thermal equilibrium at freeze-out~\cite{Andronic}.
%Since the thermal condition $(T,\mu)$ at freeze-out is extensively studied
%and verified by using several observables,
We expect the statistical model also works for the exotic hadron production,
which is calculated to be frequent enough.
%The statistical model predicts that many exotic hadrons are produced
%frequently enough to be observed.
%
One of the problems of the statistical model is its prediction power
for resonance states.
For example, the statistical model overestimates the yield of $\Lambda(1520)$
by a factor of two or more.
This discrepancy is explained based on the coalescence (recombination) model,
which uses the internal wave function
thus includes the angular momentum effects~\cite{Enyo}.
%
%The coalescence model is another successful model of hadron production.
Since the reaction time at RHIC and LHC is not very long,
it may not be reasonable to understand
the equilibrium hadron production literally
and we may need underlying hadron production mechanisms
which result in statistical distribution of ground state hadrons.

%----- FIGURE  ------------------------------------------------------- 
\begin{figure}[htb!] 
\centering 
\Fig{7.8cm}{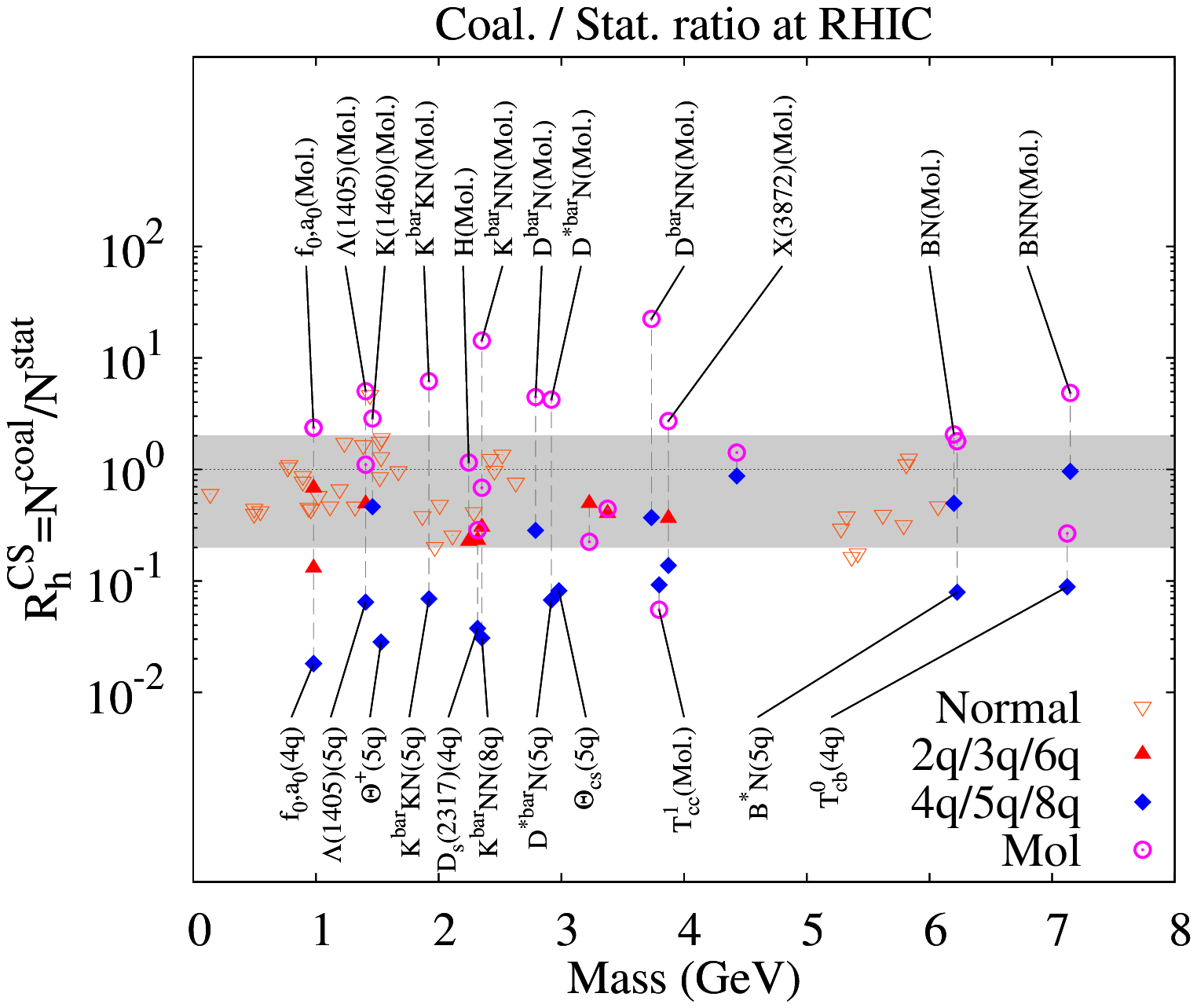}~%
\Fig{7.8cm}{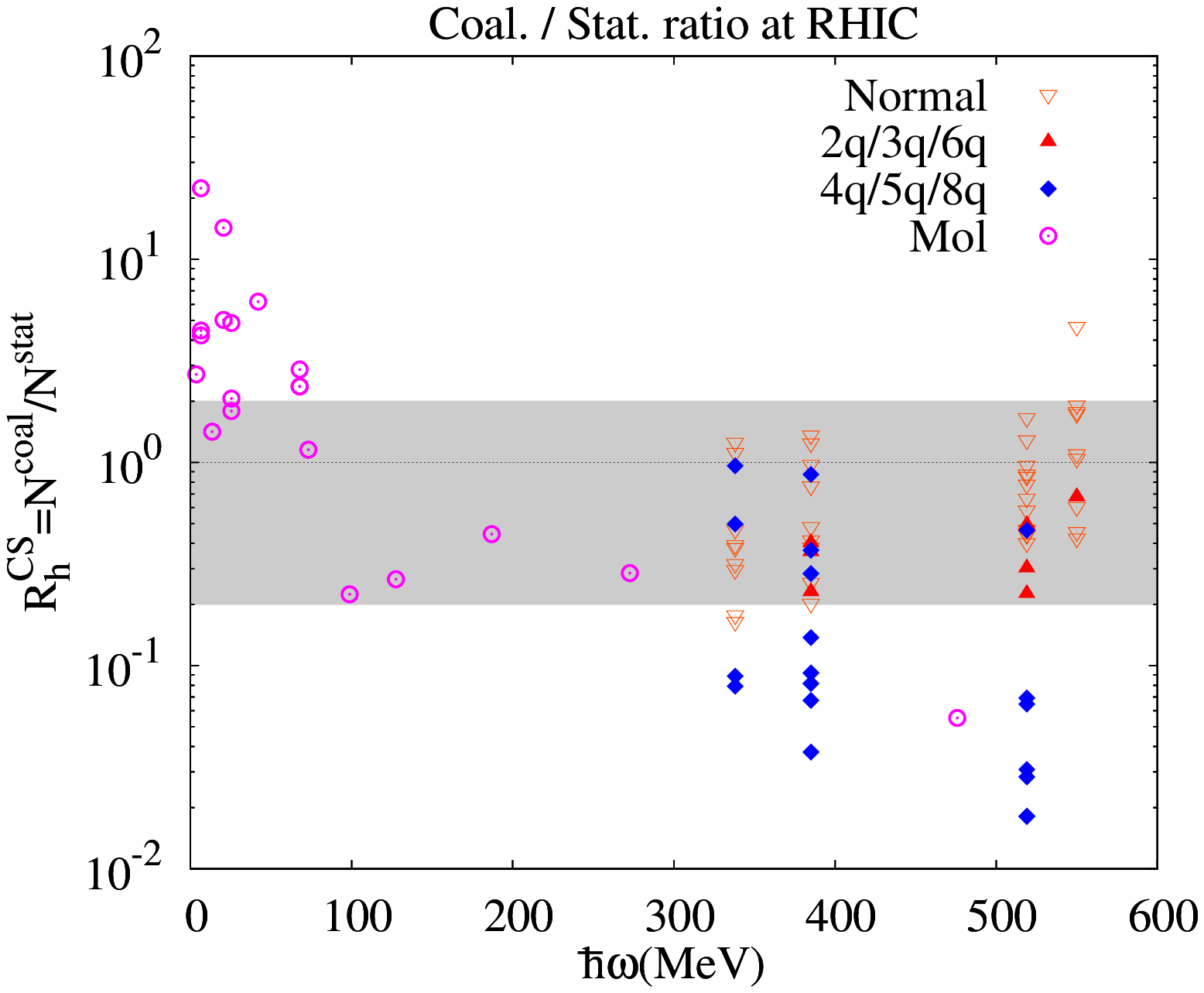}
\caption{
Coalescence / Statistical model ratio $R^\mathrm{CS}_h$
as a function of mass (left) and $\hbar\omega$ (right).
}\label{Fig:Rcs}
\end{figure} 
%--------------------------------------------------------------------- 
In order to discuss the production mechanism dependence,
we compare the results of the coalescence model
and the statistical model at RHIC and LHC.
We adopt the setup proposed by Chen and his collaborators~\cite{Chen2004}.
Hadron yields in statistical model,
\begin{align}
N_h^\mathrm{stat}=\frac{g_h V_H}{2\pi^2} \int_0^\infty 
\frac{p^2 dp}{\gamma_h^{-1} e^{(E_h-\mu_h)/T_H}\pm1}\ ,
\end{align}
are obtained at the transition (hadronization)
temperature $T=T_H=175~\MeV$ and volume $V=V_H$.
$g_h$, $\gamma_h$ and $\mu_h$ are the degeneracy, fugacity,
and chemical potential of the hadron $h$.
Light quarks ($u, d$ and $s$) are considered to reach approximate
chemical equilibrium.
Since charm and bottom quarks are produced
from initial hard scattering and their numbers are much larger than 
the equilibrium values,
we put the fugacity $\gamma_h>1$
for hadrons which contains charm and bottom quarks.
We have fixed the fugacity of hadrons containing charm and bottom quarks
to reproduce the expected number of charm and bottom quark number
from initial hard scattering.

The yield of a hadron $h$ in the coalescence model is given as,
\begin{align}
N_h^\mathrm{coal}
=&g_h \int \left[
	\prod_{i=1}^n \frac{1}{g_i} \frac{p_i\cdot d\sigma_i}{(2\pi)^3}
	\frac{d^3p_i}{E_i} f_\mathrm{th}(x_i, p_i)
	\right]
	f_W(x_1, \cdots, x_n; p_1, \cdots, p_n)
\nonumber\\
\simeq& \frac{g_hV(M\omega)^{3/2}}{(4\pi)^{3/2}}
	\frac{(2T/\omega)^L}{(1+2T/\omega)^{n+L-1}}
	\prod_{j=1}^n \frac{N_j (4\pi)^{3/2}}{g_j V(m_j\omega)^{3/2}}
	\prod_{i=1}^{n-1} \frac{(2 l_i)!!}{ (2l_i+1)!!}
\ , 
\end{align}
where
$f_\mathrm{th}$ and $f_W$ are the Wigner function of the constituents
and the intrinsic states, respectively,
$T$ and $V$ are the temperature and volume at coalescence,
$g_j$, $N_j$ and $m_j$ are the degeneracy, yield and mass
of the $j$-th constituent (hadron or quark),
$M$ is the sum of constituent masses,
and $l_i$ denotes the orbital angular momentum of the $i$-th Jacobi coordinate
and their sum is $L$.
Coalescence model calculations are performed
at the hadronization temperature $T=T_H$
and the freeze-out temperature $T=T_F$
for the quark and hadronic coalescence, respectively.
We have assumed a harmonic oscillator wave functions,
whose frequency is given as $\omega=550~\MeV$
for hadrons made of $u$ and $d$ quarks.
For hadrons containing strange, charm and bottom quarks,
we fit the statistical model results of
$\Lambda(1115)$, $\Lambda_c(2286)$ and $\Lambda_b(5620)$
and obtain $\omega_s=519~\MeV, \omega_c=385~\MeV$ and $\omega_b=338~\MeV$,
respectively.

In Fig.~\ref{Fig:Rcs}, 
we show the ratio of the hadron yields in the coalescence
and statistical models,
$R^\mathrm{CS}_h=N_h^\mathrm{coal}/N_h^\mathrm{stat}$,
where $N_h^\mathrm{coal/stat}$ denotes the hadron yields
per unit rapidity in the coalescence or statistical model.
First, we note that the ratio for normal hadrons (open triangles)
is in the range of $0.2 < R^\mathrm{CS}_h < 2$ (gray band).
%------------------------------------------------------------------------*
% Added in v2
%------------------------------------------------------------------------*
Here "normal" hadrons are defined as
particle states considered to be made of $\bar{q}q$ and $qqq$
for mesons and baryons, respectively;
lowest mass states for given quantum numbers ($J^\pi$ and flavor)
of pseudoscalar mesons ($J^\pi=0^-$),
vector mesons ($1^-$),
and $1/2^+$ and $3/2^+$ baryons.
We also categorize
$N(1440)(1/2^+), N(1520)(3/2^-), N(1535)(1/2^-)$ and $D_1(2420)(1^+)$
as normal hadrons.
%------------------------------------------------------------------------*
%
Secondly, the coalescence model is found to predict
smaller yields of compact multi-quark states.
When we use the same hadron size parameter as that of normal hadrons,
an addition of a $s$-wave, $p$-wave, or $d$-quark leads
to a suppression factor of 0.36, 0.093, or 0.029, respectively~\cite{ExHIC}.
Thus compact multi-quark states are suppressed~\cite{Nonaka}.

Another interesting feature found in the coalescence-statistical ratios
is the enhancement of spatially extended hadronic molecules.
Let us consider the two-body $s$-wave coalescence in isotropic environment.
The coalescence yield is given as the convolution of the intrinsic Wigner
function and the thermal distribution of the relative coordinate,
\begin{align}
N_h \propto \int \frac{d^Dx d^Dp}{(2\pi\hbar)^D}
 f_W(\bold{x},\bold{p}) f_\mathrm{th}(\bold{x},\bold{p}) 
=\left[
\left(\frac{4}{\hbar^2}\right)\,
\left( (\Delta p)^2+\mu T \right)
\left( (\Delta x)^2+2R^2 \right)
\right]^{-D/2}
\ ,
\end{align}
where $\Delta x$ ($\Delta p$) is the width in the intrinsic Wigner function
in the spatial (momentum) coordinate,
$\mu$ is the reduced mass, and $T$ is the temperature at coalescence.
We have assumed here the spatial Gaussian source with the radius $R$.
When the minimum uncertainty $\Delta x \Delta p=\hbar/2$ is assumed,
the above yield shows a maximum when the spatial-to-momentum width ratio
of the intrinsic Wigner function is the same
as that of the source, which reads $\hbar\omega=\sqrt{\hbar^2T/2\mu R^2}$.
As an example, in the case of $T=170~\MeV$, $\mu=500~\MeV$, $R=5~\fm$,
the optimal value of the oscillator frequency is $\hbar\omega=16~\MeV$,
which is much smaller than that of normal hadrons,
$\hbar\omega=(300-600)~\MeV$.
In the right panel of Fig.~\ref{Fig:Rcs},
we show the $\hbar\omega$ dependence of $R_h^\mathrm{CS}$.
In the present calculation,
we have assumed that the source size is large enough.
These results include three-body hadronic molecules,
but the trend is the same.
The coalescence favors hadrons whose shape in the phase space
is similar to that of the source,
then the large source size and moderate $T$ prefer extended hadrons
in coalescence.

If the coalescence is the underlying mechanism of the statistical model,
the coalescence model would give better predictions
of the hadron yields including resonances and exotic hadrons,
and we can utilize high-energy heavy-ion collisions
as a ruler of the hadron size;
smaller and larger yields
for compact multi-quark states and spatially extended hadronic molecule states,
respectively.

\section{Exotic Interaction from heavy ion collisions
--- $\Lambda\Lambda$ interaction ---}
\label{Sec:LL}

Where is the $S=-2$ dibaryon, $H$ ?
This is a long standing problem in hadron physics.
In 1977, Jaffe pointed out that double strange dibaryon 
made of 6 quarks ($uuddss$) may be deeply bound below the $\Lambda\Lambda$
threshold due to the strong attraction
from color magnetic interaction~\cite{Jaffe}.
%The predicted binding energy is around 80 MeV.
%
Dedicated experiments have been performed to find the $H$ particle
in these 35 years.
Deeply bound $H$ was denied by the observation of double $\Lambda$ hypernuclei.
For example, a double $\Lambda$ hypernucleus
$^{~~6}_{\Lambda\Lambda}\mathrm{He}$ 
was found to decay weakly in the Nagara event,
and the observed energy of $^{~~6}_{\Lambda\Lambda}\mathrm{He}$
is $6.91~\MeV(=B_{\Lambda\Lambda})$
below the $^4\mathrm{He}+\Lambda\Lambda$ threshold~\cite{Nagara}.
If the mass of $H$ is below $2M_\Lambda-B_{\Lambda\Lambda}$,
$^{~~6}_{\Lambda\Lambda}\mathrm{He}$ should decay to
$^4\mathrm{He}+H$ strongly.
%
%Why does not the H deeply bound ?
The reason why the attraction is weaker than expected
%Weaker attraction than expected 
may be the determinant-type 
3-quark interaction~\cite{KMT}, which is repulsive 
in the $H$ channel~\cite{OkaTakeuchi}.
While the deeply bound $H$ is denied,
the attraction in the $H$ channel may generate a pole
in the weakly bound or resonance region.
There are some hints in recent experimental and theoretical studies.
The KEK-E522 experiment observed
a bump in the $\Lambda\Lambda$ invariant mass spectrum~\cite{E522}.
Recent lattice QCD studies imply that $H$ should exist as a bound state
in the SU(3) limit and/or with heavy pion masses~\cite{Lattice}.
Thus the physics of the $H$ particle is
a long-standing as well as current problem.

Existence of the $H$ particle state is
closely related to the $\Lambda\Lambda$ interaction.
The $H$ particle pole is, if exists, near the \LL\ threshold,
and it is natural to expect that
%$H$ is a bound state or a resonance state containing a significant component of \LL.
$H$ contains a significant component of \LL.
\LL\ interaction is important also for the dense matter equation of state (EOS).
In many of theoretical calculations,
$\Lambda$ fraction in dense neutron star matter is
compatible with the neutron fraction,
then the strength of the \LL\ interaction may affect the EOS.
Until now, available information on \LL\ interaction is scarce.
We know that it is weakly attractive from the \LL\ bond energy
in $^{~~6}_{\Lambda\Lambda}\mathrm{He}$,
$\DBLL=
B_{\Lambda\Lambda}(^{~~6}_{\Lambda\Lambda}\mathrm{He})
-2 B_{\Lambda}(^{5}_{\Lambda}\mathrm{He}) \simeq 0.6~\mathrm{MeV}$.
From $\DBLL(^{~~6}_{\Lambda\Lambda}\mathrm{He})$,
the scattering length and the effective range 
in the \LL\ $^1\mathrm{S}_0$ channel 
are obtained as
%-------------------------------------------------------------*
% Corrected in v2: FG and Hiyama values are exchanged to be correct.
%-------------------------------------------------------------*
$(a_0, \reff)=(-0.77~\fm, 6.59~\fm)$~\cite{FG}
or 
$(a_0, \reff)=(-0.575~\fm, 6.45~\fm)$~\cite{Hiyama},
%-------------------------------------------------------------*
but in principle we cannot determine two low energy scattering parameters
from one observed value of $\DBLL$.
%$a_0=-0.77~\fm, \reff = 6.59~\fm$ (Filikhin, Gal)
%$a_0=-0.575~\fm, \reff = 6.45~\fm$ (Hiyama et al.)
%These values are obtained
%by tuning the Nijmegen model parameters to fit $\DBLL$.

Thus other observational information on \LL\ interaction has been desired.
One of the ways is to observe the binding energies
of various double $\Lambda$ hypernuclei,
as planned in the J-PARC E07 experiment.
Another available observable is the \LL\ correlation in nuclear reactions.
Actually, KEK-E522 experiment~\cite{E522} has demonstrated that 
\LL\ invariant mass spectrum is enhanced in the low energy region
compared with the phase space estimate
and the classical transport model calculation,
implying that \LL\ interaction is attractive.
In high-energy heavy-ion collisions,
abundant $\Lambda$ particles are produced,
and we can measure the \LL\ relative momentum correlation,
which contains information on \LL\ interaction.
This idea is not new.
It was proposed in '80s that we can fix resonance parameters,
when the source size is small~\cite{GreinerMuller1989}.
The correlation at low relative momenta was proposed to be useful
to discriminate the sign of the scattering length $a_0$,
provided that the source size is large~\cite{OHNSA};
When \LL\ has a bound state ($a_0>0$),
the scattering wave function must have a node
at $r \simeq a_0$
in order to be orthogonal to the bound state wave function,
then we may find the suppression of the correlation.
Now RHIC and LHC have the vertex detectors,
and we can really obtain the \LL\ correlation data in heavy-ion collisions.

%\subsection{\LL\ correlation in heavy-ion collisions and \LL\ interaction}

We here discuss \LL\ correlation in heavy ion collisions
in view of \LL\ interaction.
Two particle correlation at low relative momentum
from a chaotic source is known to be sensitive to the source size
and the two particle interaction~\cite{Bauer1992}.
The \LL\ correlation function is given as,
\begin{align}
C_{\Lambda\Lambda}(q)
=& \frac{
	\int d\bold{x}_1 d\bold{x}_2
	S(\bold{x}_1,\bold{p}+\bold{q})
	S(\bold{x}_2,\bold{p}-\bold{q})
	\left| \psi^{(-)}(\bold{x}_{12},\bold{q}) \right|^2
  }{
	\int d\bold{x}_1 d\bold{x}_2
	S(\bold{x}_1,\bold{p}+\bold{q})
	S(\bold{x}_2,\bold{p}-\bold{q})
  }
\nonumber\\
\simeq & 1 - \frac12 \exp(-4q^2 R^2) + \frac12 \int d\bold{r} S_{12}(\bold{r})
	\left(
		 \left| \chi_0(\bold{r})\right|^2
		-\left| j_0(qr)\right|^2
	\right)
\ ,
\end{align}
where
$\psi^{(-)}(\bold{x},\bold{q})$ is the relative wave function
having the relative momentum $\bold{q}$ in the final state,
$\chi_0$ is the relative wave function in the $s-$wave,
and $S$ denotes the source function.
In obtaining the second line,
we have made following two approximations.
(1) The single particle source function has a Gaussian profile
whose width is independent of momentum,
then the source function in the relative coordinate is given as
$S_{12}(\bold{r})=(2R\sqrt{\pi})^{-3}\,\exp(-r^2/4R^2)$.
%We discuss the flow effects later.
(2) Only the $s-$wave relative wave function is modified
from the free case by the \LL\ interaction.

%----- FIGURE  ------------------------------------------------------- 
\begin{figure}[htb!] 
\centering 
\Fig{7.8cm}{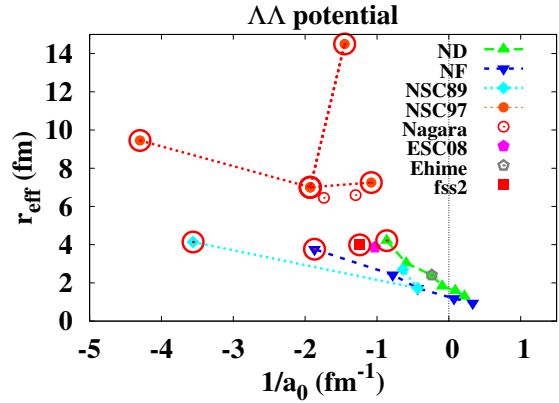}
\caption{
Scattering parameters from \LL\ interaction models.
}\label{Fig:LLpotpars}
\end{figure} 
%--------------------------------------------------------------------- 

The correlation function is determined by the source function
and the relative wave function, the latter of which is sensitive
to the \LL\ interaction.
We compare the results with several types of \LL\ interactions.
The first type of \LL\ interactions is the Nijmegen models~\cite{Nijmegen},
which are based on a meson and meson-pair exchange picture 
of baryon-baryon interactions.
The second type of baryon-baryon interaction is
the quark model interaction fss2~\cite{fss2},
which takes account of the Pauli blocking at the quark level,
gluon exchanges between quarks, and meson exchanges.
We also compare the results of a one-boson exchange \LL\ interaction,
Ehime potential~\cite{Ehime}.
Since the Ehime potential is proposed before the Nagara event,
it assumes a smaller \LL\ bond energy,
$\DBLL(^{~~6}_{\Lambda\Lambda}\mathrm{He})=3.6~\MeV$,
than that obtained in the Nagara event.
In actual calculations, we use two range Gaussian potentials
which fit the scattering length and effective range
for Nijmegen and Ehime potentials.
For fss2, we use a phase-shift equivalent local potential,
derived by using the inversion method
based on supersymmetric quantum mechanics~\cite{fss2,SB}.
Scattering parameters $(a_0, \reff)$ of these interactions
are shown in Fig.~\ref{Fig:LLpotpars}.

%----- FIGURE  ------------------------------------------------------- 
\begin{figure}[htb!] 
\centering 
\Fig{7.8cm}{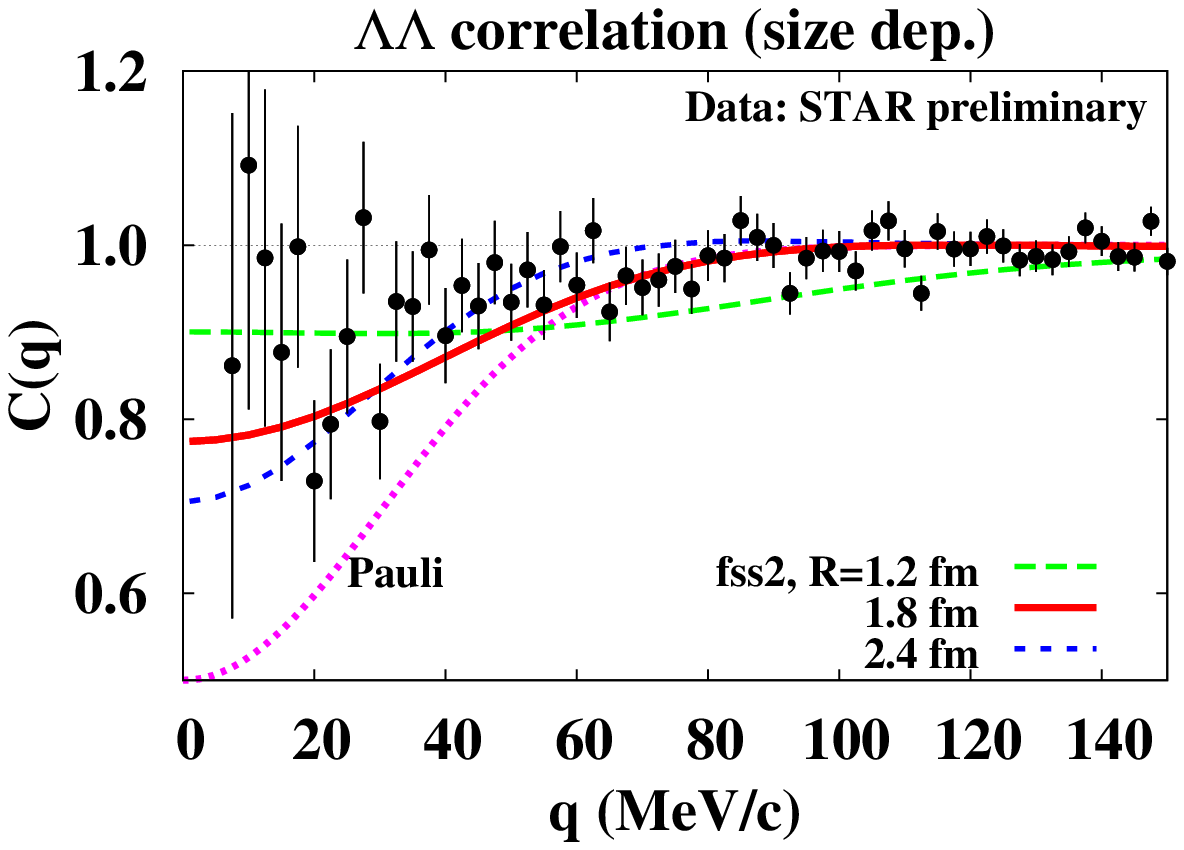}%
~\Fig{7.8cm}{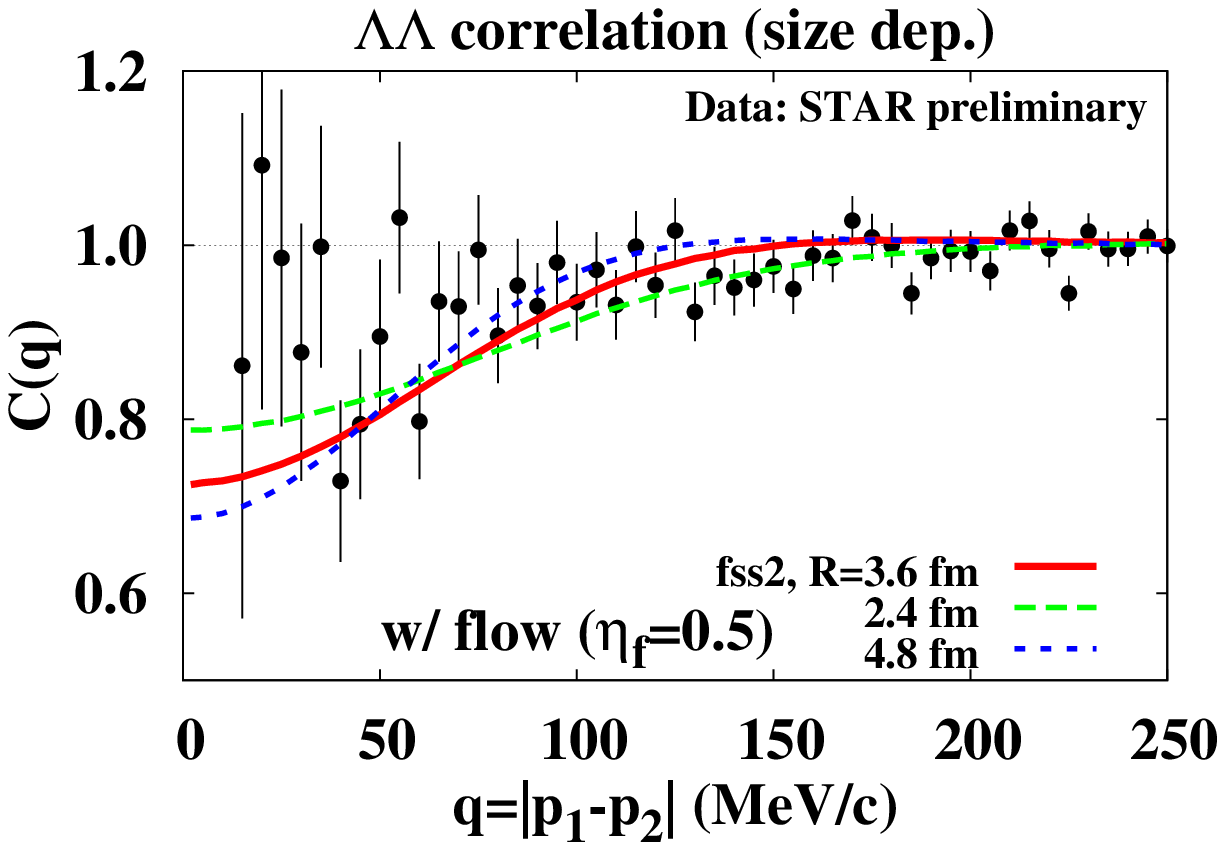}%
\caption{
Left: Source size dependence of the \LL\ correlation.
Right: Calculated and observed \LL\ correlation.
}\label{Fig:size}
\end{figure} 
%--------------------------------------------------------------------- 
Now we shall try to extract the information on the \LL\ interaction
as well as the source size from the \LL\ correlation data.
%While the error bars are still large,
%the observed data provides precious information
%to constrain \LL interaction.
In the left panel of Fig.~\ref{Fig:size},
we show the source size dependence of the \LL\ correlation $C(q)$.
Calculated results using fss2, as an example, are compared
with the RHIC-STAR data~\cite{STAR}.
When the source size is small,
$C(q)-1$ has a long tail in the high momentum region.
Thus we can obtain the source size by fitting the correlation  
at high momenta.
The optimal source radius $R$ depends on the \LL\ interaction,
but the data suggest that the apparent source size of $\Lambda$
is around $R \sim 2~\fm$,
which is smaller than the pion and kaon source.
This difference may be due to the flow effects.
In the right panel of Fig.~\ref{Fig:size},
we show the size dependence of the \LL\ correlation
with flow effects.
Transverse flow generally extends the range of finite correlation 
to a higher momentum region.
Extending the correlation to higher momentum means that
the flow makes the apparent radius smaller than the actual source size.
We find that the $\Lambda$ source size may be in the range 
$3~\fm \lesssim R \lesssim 4~\fm$
when we take a reasonable flow parameter $\eta_f=0.5$
(the transverse rapidity is given as $Y_T=\eta_f r_T/R$).
This source size would be consistent with the proton source size.

\begin{figure}[htb!] 
\centering 
\Fig{7.8cm}{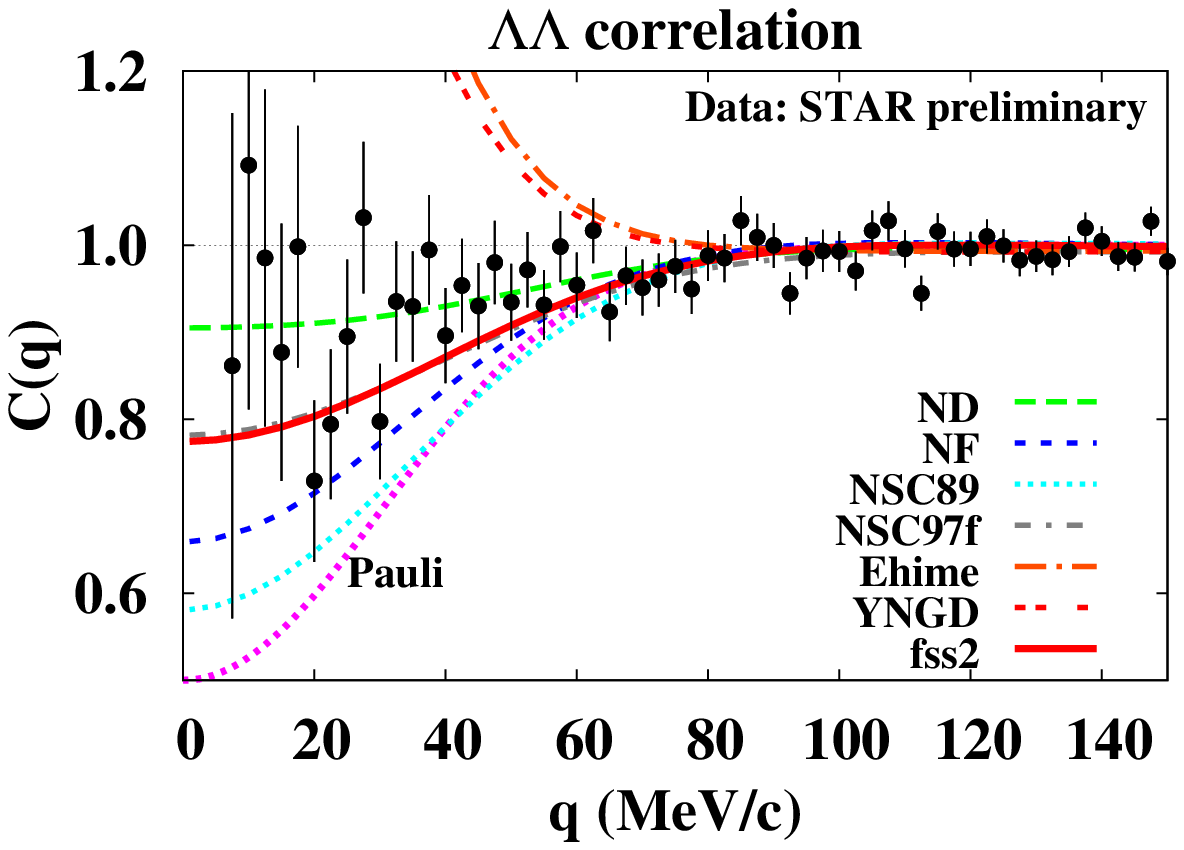}%
~\Fig{7.8cm}{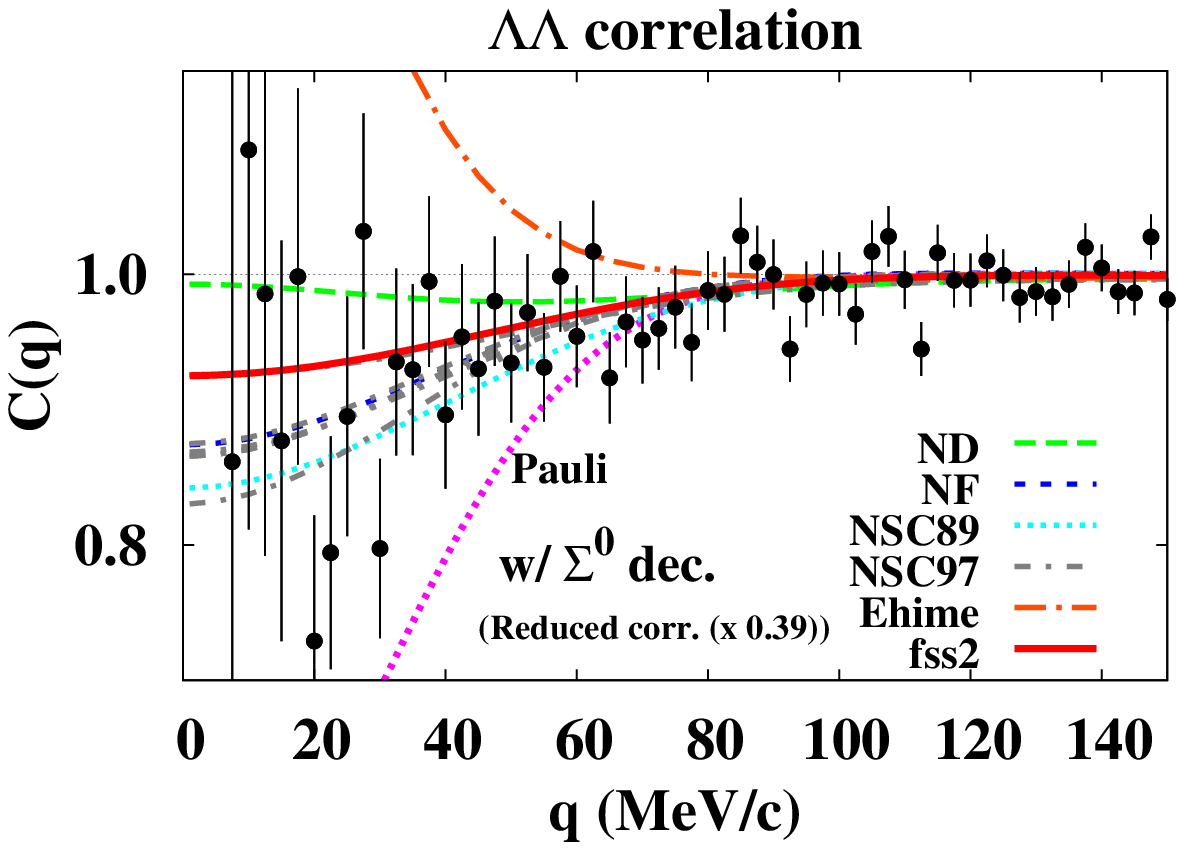}
\caption{
\LL\ correlation obtained by using several \LL\ interactions.
The source size is chosen to fit the high momentum tail region
of the correlation.
Left (right) panel shows the results without (with) $\Sigma^0$ decay effects.
}\label{Fig:HLL}
\end{figure}

After fitting the tail region,
we can discriminate the \LL\ interaction
from the behavior of $C(q)$ at small $q$.
In the left panel of Fig.~\ref{Fig:HLL},
we compare the results from several \LL\ interactions.
It seems that Nijmegen model D (ND)
with the hard core radius of $R_c=0.56~\fm$,
Nijmegen soft-core 97 model (NSC97f),
and quark model interaction (fss2)
are consistent with the RHIC-STAR data.
It should be noted that
these results are obtained in a simple setup;
the single channel calculation,
no feeddown effects,
and no flow effects.

%----- FIGURE  ------------------------------------------------------- 
%--------------------------------------------------------------------- 
For quantitative discussions,
we need to consider the feeddown effects from heavier particles.
The feeddown effects have been known to be decisive for the $pp$ correlation.
The $pp$ correlation at low momentum
is suppressed by the Coulomb repulsion,
and the Gamow factor correction recovers $C(q)$ to be around unity.
In high-energy heavy-ion collisions,
we have protons also from the decay of heavier particles, 
such as $\Lambda \to p\pi^-$.
There is no Coulomb suppression in the $p\Lambda$ channel,
and the pionic weak decay does not change the baryon momentum much.
As a result, $p\Lambda$ correlation in the reaction region
strongly affects the $pp$ correlation.
Compared with the $pp$ correlation,
the feeddown effects on the \LL\ correlation are expected to be small.
There is no Coulomb suppression in the \LL\ channel,
and the particles which decays into $\Lambda$ are limited.
Furthermore, 
it is possible to exclude the $\Lambda$ from weak decay
such as $\Xi^-\to\Lambda\pi^-$ using the vertex detectors, 
if necessary.
There exists an exception, $\Sigma^0\to\Lambda\gamma$,
which we cannot exclude experimentally.
It is not easy to detect $\gamma$ decay vertex,
then we should take $\Sigma^0$ decay effects 
in theoretical estimates.
%The left panel of Fig.~\ref{Fig:wDecay} shows the $\Sigma^0$ decay
%effects on the \LL\ correlation.
We find that we can simulate the decay effects 
by multiplying a factor 0.39 to $C(q)-1$, 
if the pre-decay correlation in the $\Sigma\Lambda$ or $\Sigma\Sigma$
channel is small.
%The gray zone shows the range of \LL\ correlation
%when $C(q)-1=\pm 0.1 (q\to0)$ is assumed in the pre-decay channels.
For more serious studies, 
we need to take account of the $\Sigma\Lambda$ and $\Sigma\Sigma$
interaction and correlation in a given model of $BB$ interaction consistently.

In the right panel of Fig.~\ref{Fig:HLL},
we compare \LL\ correlation using \LL\ interactions
under consideration.
We have included the $\Sigma^0$ decay effects
by the above mentioned simple prescription.
We have not made the $\chi^2$ analysis,
but the \LL\ correlation data seems to favor
fss2, NF($R_c=0.50~\fm$) and some versions of NSC97 interactions.
ND($R_c=0.56~\fm$) and NSC89($m_\mathrm{cut}=1020~\MeV$)
may be also allowed.
These favored interactions are shown in the $(1/a_0, \reff)$ plane
marked with open circles in Fig.~\ref{Fig:LLpotpars}.
We conclude that \LL\ interactions with
$1/a_0 \leq -0.8~\fm^{-1}$ and $\reff \geq 3~\fm$
are favored by the recent \LL\ correlation preliminary data
in high-energy heavy-ion collisions at RHIC
by STAR collaboration~\cite{STAR}.
These results are consistent with the analysis of the Nagara event,
which is based on NSC97 interactions~\cite{FG,Hiyama}.
A more recent Nijmegen interaction, ESC08~\cite{ESC08},
has a similar scattering parameters to fss2,
and it is also in the above mentioned range.

%----- FIGURE  ------------------------------------------------------- 
\begin{figure}[htb!] 
\centering 
\Fig{7.8cm}{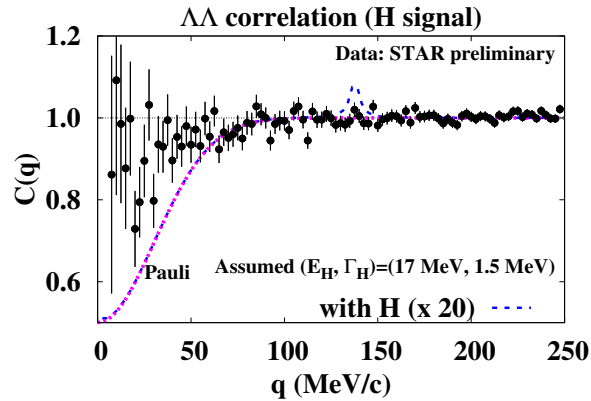}%
\caption{
$H$ signal in \LL\ correlation.
The yield of $H$ is assumed to be 20 times larger
than the statistical model result.
}\label{Fig:H}
\end{figure} 
%--------------------------------------------------------------------- 
There are some more effects to be discussed
for a more quantitative discussion on \LL\ interactions.
Another one is the coupling effects, $\Lambda\Lambda \leftrightarrow \Xi N$.
Here we have included the coupling effects in a simple coupling potential,
$V_{\Lambda\Lambda-\Xi N}(r)=v_\mathrm{cpl} \exp(-r^2/b^2)$ with $b=1~\fm$.
When the coupling is strong,
the \LL\ component is suppressed in the inner part
of the relative wave functions.
As a result, the coupling potential acts as repulsive interaction.
We find that the coupling effects to $\Xi N$ channel are mild,
as long as the coupling potential is not very strong,
$v_\mathrm{cpl}<50~\MeV$.

Finally, we discuss the existence of the $H$ particle.
We assume that the mass of $H$ is 17 MeV above the \LL\ threshold,
%-------------------------------------------------------------*
% Corrected in v2: 15 MeV --> 1.5 MeV
%-------------------------------------------------------------*
and its width is 1.5 MeV.
%-------------------------------------------------------------*
For the yield, the statistical model result is adopted as a reference value.
Figure~\ref{Fig:H} shows the $H$ signal which would be observed
in the \LL\ correlation, but with 20 times larger yield than
the statistical model result.
At present, we do not see any signal of $H$, 
but we need much more precision to conclude the existence of $H$
from \LL\ correlation unfortunately.
This is because the number of background \LL\ pair is large,
then the signal-to-noise ratio is small for the $H$ above the threshold.

\section{Summary}
\label{Sec:Summary}

We have discussed the exotic hadron size and hadron-hadron interactions
in terms of the hadron yield and hadron-hadron correlation
in heavy ion collisions.

In the first part,
we have demonstrated that heavy ion collisions may play a role of
the hadron size ruler:
In the framework of the coalescence model,
a hadron with a large size would be produced more abundantly
compared with the statistical model result.
The mechanism of this enhancement is argued
in a simple two body coalescence case, 
and we have found that coalescence favors hadrons
whose shape in the phase space is similar to that of the source.
If this coalescence mechanism also applies to other reactions,
the yield of compact multi-quark states may be larger
than the statistical model result in $e^+e^-$ reaction.
Thus it will be valuable to investigate exotic hadron production
in $e^+e^-$ collisions.
We have assumed that the source size is large enough in heavy-ion collisions,
but it is necessary to take account of the finite size effects
in $e^+e^-$, where the source size would be compatible with the hadron size.

In the second part,
we have discussed the \LL\ interaction
and its effects on the \LL\ correlation,
which is recently measured at RHIC~\cite{STAR}.
Based on a Gaussian source assumption
and by considering the decay effects of $\Sigma^0\to\Lambda\gamma$,
we have compared the data with calculated results using several types
of \LL\ interaction.
We find that the RHIC-STAR data favor 
the \LL\ scattering parameters in the range
$1/a_0 \leq -0.8~\fm^{-1}$ and $\reff \geq 3~\fm$.
These are consistent with the \LL\ interaction parameters
which reproduce the \LL\ bond energy in $^{~~6}_{\Lambda\Lambda}\mathrm{He}$
and recent \LL\ interactions.
%------------------------------------------------------------------------*
% Added in v2
%------------------------------------------------------------------------*
The \LL\ correlation data at low relative momenta
%from the RHIC-STAR collaboration
seem to unfavor the existence of
the bound $H$ state containing significant \LL\ component.
In order to identify/rule out the resonance $H$ state
by using \LL\ correlation data at higher relative momenta,
we need more statistics because of the large continuum \LL\ pair yield.
%------------------------------------------------------------------------*
For more serious estimate of the \LL\ interaction,
it is necessary to combine the feeddown, flow, and couple channel effects
simultaneously.

\section{Acknowledgments}
%This work is supported by ....
%\medskip
The authors would like to thank A. Gal, J. Schaffner-Bielich, J. Aichelin,
and A. Sandorfi for useful suggestions.
This work is supported in part by the Grants-in-Aid for Scientific Research
from JSPS
(Nos.
          (B) 23340067, %(T. Kunihiro (incl. A.Ohnishi))
          (B) 24340054, %(A. Nakamura (incl. A.Ohnishi))
          (C) 24540271  %(A. Ohnishi, K. Morita, T. Kunihiro)
),
by the Grants-in-Aid for Scientific Research on Innovative Areas from MEXT
(No. 2404: 24105001, 24105008), % AO (NS: X01, D01)
by the Yukawa International Program for Quark-hadron Sciences,
and by the Grant-in-Aid for the global COE program ``The Next Generation
of Physics, Spun from Universality and Emergence" from MEXT.



%% References


%\bibliographystyle{elsarticle-num}



\end{document}